\documentclass[journal]{IEEEtran}
\ifCLASSINFOpdf
\else
\fi
\usepackage{graphicx}
\usepackage{amsmath}
\usepackage{upgreek}
\usepackage{cite}

\begin{document}
%
\title{Early-Stage Cancer Biomarker Detection via Intravascular Nanomachines: Modeling and Analysis}
%
%
%

\author{Abdollah~Rezagholi,
        Sergi~Abadal,~\IEEEmembership{Senior~Member,~IEEE},
        Filip~Lemic,
        Eduard~Alarcón,~\IEEEmembership{Member,~IEEE},
        and~Ethungshan~Shitiri 
\thanks{Abdollah~Rezagholi, Sergi~Abadal, Eduard~Alarcón, and Ethungshan~Shitiri are with Nanonetworking Center in Catalunya (N3Cat), Universitat Politècnica de Catalunya (UPC), Spain. e-mail: \{abdollah.rezagholi, sergi.abadal, eduard.alarcon, ethungshan.shitiri\}@upc.edu.}%
\thanks{Filip~Lemic is with the i2Cat Foundation, Spain. e-mail: filip.lemic@i2cat.net.}%
\thanks{This project has received funding from the EU's Horizon Europe research and innovation programme under the Marie Skłodowska-Curie grant agreement No 101154851.}
}

\maketitle

\begin{abstract}
Early detection of cancer is essential for timely diagnosis and improved patient outcomes. Among emerging technologies, intra-body nanoscale communication offers an innovative solution to identify molecular cues within the human bloodstream. This study investigates a minimally invasive approach for early-stage cancer biomarker detection using nanomachines introduced into the bloodstream. To assess the feasibility of this approach, computational simulations are used to emulate the vascular environment and evaluate biomarker detection performance under different physiological conditions. Current modeling approaches often fail to capture essential vascular characteristics, including non-uniform flow structures, size-dependent particle mobility, and particle margination driven by red blood cell interactions. To address these limitations, our study incorporates these factors into the simulation framework and quantifies their individual and combined effects on biomarker detection efficiency. Baseline detection performance is first obtained under uniform flow assumptions, after which introducing realistic vascular transport mechanisms progressively reduces detection probability for all vessel types and nanomachine sizes. Among the considered vessels, capillary consistently achieves the highest detection probability across all nanomachine sizes.
\end{abstract}


%
\IEEEpeerreviewmaketitle

\section{Introduction}
%
%
%
%
\IEEEPARstart{E}{arly-stage} cancer detection relies on identifying disease-associated signals before clinical symptoms emerge, a goal that is fundamentally constrained by the scarcity of tumor-derived biomarkers in peripheral blood during early-stage disease \cite{tarro2005early}. Established circulating biomarkers include circulating tumor DNA (ctDNA), circulating tumor cells (CTCs), extracellular vesicles such as exosomes, and selected proteins and microRNAs \cite{pulumati2023technological}. Among these, ctDNA is currently the most clinically validated for molecular profiling and disease monitoring. However, in early-stage cancer, ctDNA typically constitutes a minute fraction of total cell-free DNA and is highly fragmented and short-lived, leading to severe sensitivity and reproducibility challenges for blood-based assays \cite{yu2024multi}. Similar limitations apply to CTCs and exosomes, whose abundance and heterogeneity further complicate reliable detection at early stages \cite{wan2017liquid}. As a result, conventional blood tests and even state-of-the-art molecular assays struggle to achieve consistent early detection without extensive sample processing and amplification.

\begin{figure}[!t]
  \centering
   \includegraphics[trim= 2.3cm 4.8cm 3.7cm 4.5cm,clip,width=\columnwidth]{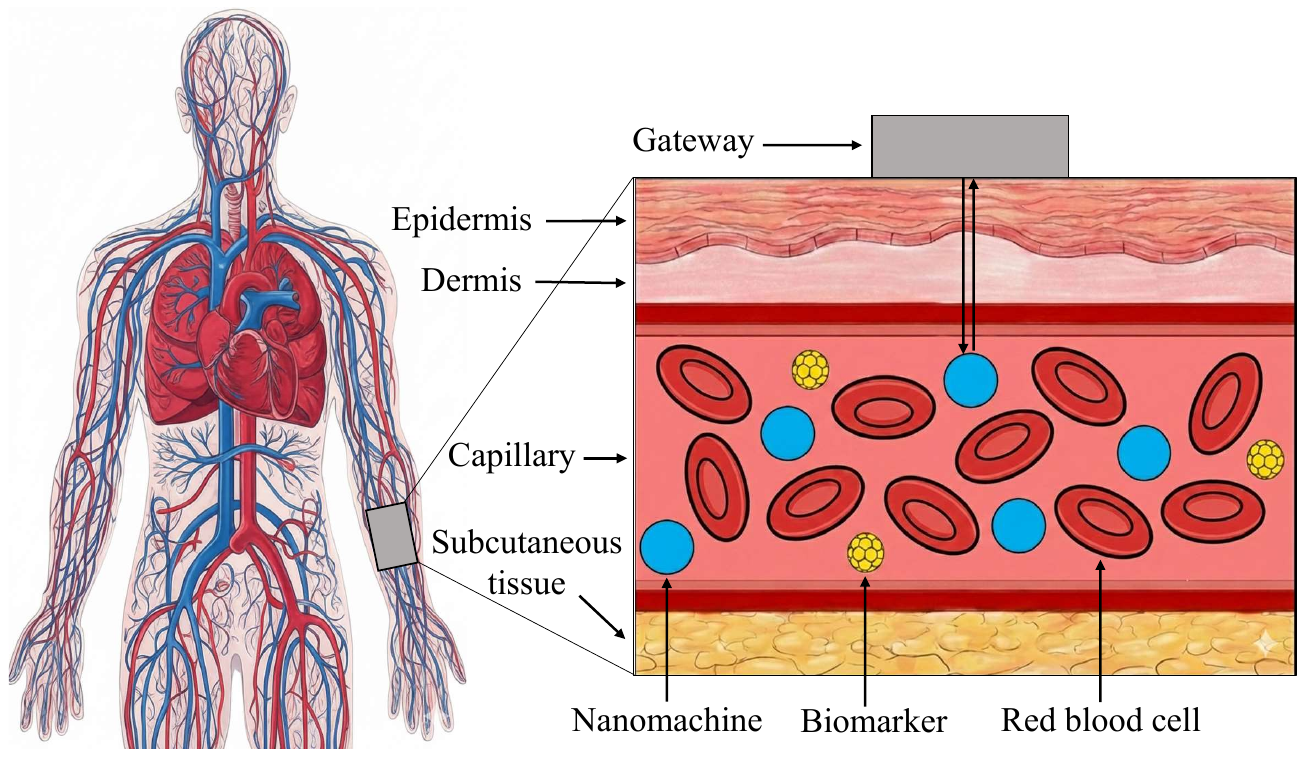}
  \caption{Schematic representation of an intra-body nanonetworks for in-vivo nanosensing and connection to external world.}
  \label{fig:Nano_Network1}
\end{figure}

To address these limitations, liquid biopsy has emerged as a minimally invasive diagnostic paradigm that enables longitudinal access to circulating biomarkers through blood or plasma samples \cite{wan2017liquid, wang2024research, chen2019next}. Within this framework, circulating exosomes provide a representative class of biomarkers accessible through such sampling approaches. In particular, ALPPL2$^{+}$ exosomes have been reported as indicative of early pancreatic abnormalities, with concentrations increasing from the order of $10^{7}\,\mathrm{mL}^{-1}$ in healthy controls to $10^{8}\,\mathrm{mL}^{-1}$ in non-cancerous conditions (NCC), and further to the $10^{9}\,\mathrm{mL}^{-1}$ range in pancreatic cancer patients~\cite{halder2025exosomal}. The NCC regime reflects the earliest measurable deviation from healthy baseline and indicates the presence of preliminary biological abnormalities. Despite these measurable changes, in current clinical practice, liquid biopsy is primarily used for tumor genotyping, therapy selection, and disease monitoring rather than population-level early detection, reflecting persistent biological and technical constraints \cite{corcoran2018application}.

In parallel, biosensor technologies, including microfluidic and nanotechnology-enabled platforms, have been extensively investigated at the research and preclinical level as potential tools to improve sensitivity, reduce assay time, and integrate sample handling with detection \cite{liu2024microfluidic, ramesh2022nanotechnology}. These approaches achieve high analytical performance under controlled conditions by operating in episodic, ex-vivo measurement settings, where detection depends on capturing transient circulating biomarkers present at low concentrations during early-stage disease.

Despite these advancements, detecting cancer at its earliest stages still remains challenging because of the inherently low quantity and instability of circulating biomarkers, highlighting the need for innovative, minimally invasive diagnostic approaches capable of continuous molecular monitoring  \cite{wahab2023biomarkers}. In-vivo sensing has been also proposed recently, which partially solves these problems by introducing the sensors into a blood vessel with an implanted needle \cite{zhu2022real}. However, this approach is invasive and is only able to localize biomarkers that pass by the vessels that are in touch with the sensor, which reduces the probability of detection.

In this context, nanotechnology opens a window of opportunity through the creation of advanced nanosensors and nanomachines. Nanosensors stand out for their remarkable accuracy in event detection \cite{ma2025nano}, while nanomachines are projected to perform basic functions like computing, communication, and actuation \cite{elayan2017bio, akyildiz2011nanonetworks} at very small scales, and could be used for the monitoring and detection of cancer. In particular, the combination of in-vivo sensing and interconnected nanomachines leads to concept of intra-body nanonetworks (IBNs), which could possess the capability of detecting cancer biomarkers directly in the bloodstream without the need for implanted needles \cite{elayan2017bio, shubair2015vivo}.

Generally, IBNs refer to systems of nanomachines that can operate within various parts of the body, such as the bloodstream, gastrointestinal tract, brain, or even within specialized implants with multiple applications \cite{moser2024liquid, cheng2023advances, kong2023advances}. In the context of this work, we propose the use of IBNs for in-vivo nanosensing for early-stage cancer detection. Fig.~\ref{fig:Nano_Network1} illustrates the proposed system, where nanomachines flow passively within blood vessels to detect biomarkers, including those from cancer tissue. Since the nanomachines flow through the circulatory system, they could flow close to the tumor that is releasing the biomarkers, increasing the detection probability. However, due to their small size, individual nanomachines may not able to notify their detection to the external world. To remedy this, nanomachines would need extra methods to communicate this information to other nearby nanomachines and, collectively, connect wirelessly to on-body gateways that bridge to external networks and enable real-time disease tracking and diagnostics.

Several studies have explored wireless communication-based approaches within the human cardiovascular system. For instance, the work from \cite{gomez2022nanosensor} employs machine learning models to estimate sensor distributions and localize abnormalities within the entire circulatory system. Another work investigates in-vivo anomaly detection by tracking the trajectories of freely injected bionanosensors using inertial positioning and backscattering communication \cite{simonjan2021body}. Although these frameworks demonstrate the feasibility of in-vivo anomaly detection, they omit key intra-vascular transport phenomena. Specifically, they do not account for laminar flow, margination, or size-dependent effects that might significantly impact on the system performance as illustrated later in this paper.

In parallel to these detection efforts, molecular communication (MC) offers a theoretical framework for modeling nanoscale interactions within the bloodstream \cite{felicetti2025molecular}. MC treats molecule propagation and reception as components of a communication channel and requires particle transport models to analyze their performance. To that end, works typically use diffusion and advection models \cite{hofmann2024molecular, gomez2024dna, scazzoli2024molecular, kianfar2024wireless, cheng2024channel}. A more in depth study of the transport effects on MC channels is given in \cite{yue2024bio}, where differences in propagation between different types of vessels, or the role of laminar flow are assessed. Again, however, the present studies neglect intra-vascular transport phenomena such as margination and size-dependent movement of particles.

In this paper, we aim to bridge the existing gap in the literature of IBN and MC with two main contributions. On the one hand, we provide a microvascular simulation model that incorporates physiologically realistic blood flow across multiple vessel types and particle sizes, hence establishing accurate transport conditions for biomarker detection. On the other hand, we use the proposed model to assess the impact of currently ignored transport phenomena in the particular case of biomarker sensing for early-stage cancer detection applications. In particular, we set the focus on three key intra-vascular phenomena: (1) laminar flow, which defines the parabolic velocity profile across the vessel \cite{hofmann2024molecular}; (2) margination, driven by red blood cell interactions that push larger particles toward the vessel walls \cite{liu2018nanoparticle}; and (3) size-dependent effects, where variations in particle size govern mobility and localization within the flow \cite{verkhovskii2021effect}. 

As part of our contribution, we implement the abovementioned mechanisms in AcCoRD \cite{noel2017simulating}, a MC simulator widely used in the nanonetworks community. This allows quantitative evaluation of their individual and combined influence on MC systems in general and, in the context of this paper, biomarker detection in microvascular environments in particular. Our study shows that, across all three considered vessel types, namely capillaries, venules, and arterioles, the adoption of laminar flow reduces detection probability due to diminished near-wall particle velocities. Higher margination levels further diminishes detection efficiency by increasing nanomachine residence near vessel boundaries, reducing interaction with freely circulating biomarkers. In contrast, increasing nanomachine size improves detection efficiency owing to the enlarged effective interaction area, despite experiencing stronger margination effects.

The remainder of this paper is organized as follows. Section~\ref{sec:model} presents the system model, including the key intra-vascular effects. Section~\ref{sec:methodology} describes the simulation methodology, detailing the simulation framework and parameter settings. Section~\ref{sec:results} shows the simulation results, illustrating how variations in vascular parameters and particle properties influence detection efficiency. Finally, Section~\ref{sec:conclusion} concludes the paper.

\begin{figure*}[!t]
\centering
\includegraphics[trim=0 5.77cm 0 6.8cm,clip,width=\textwidth]{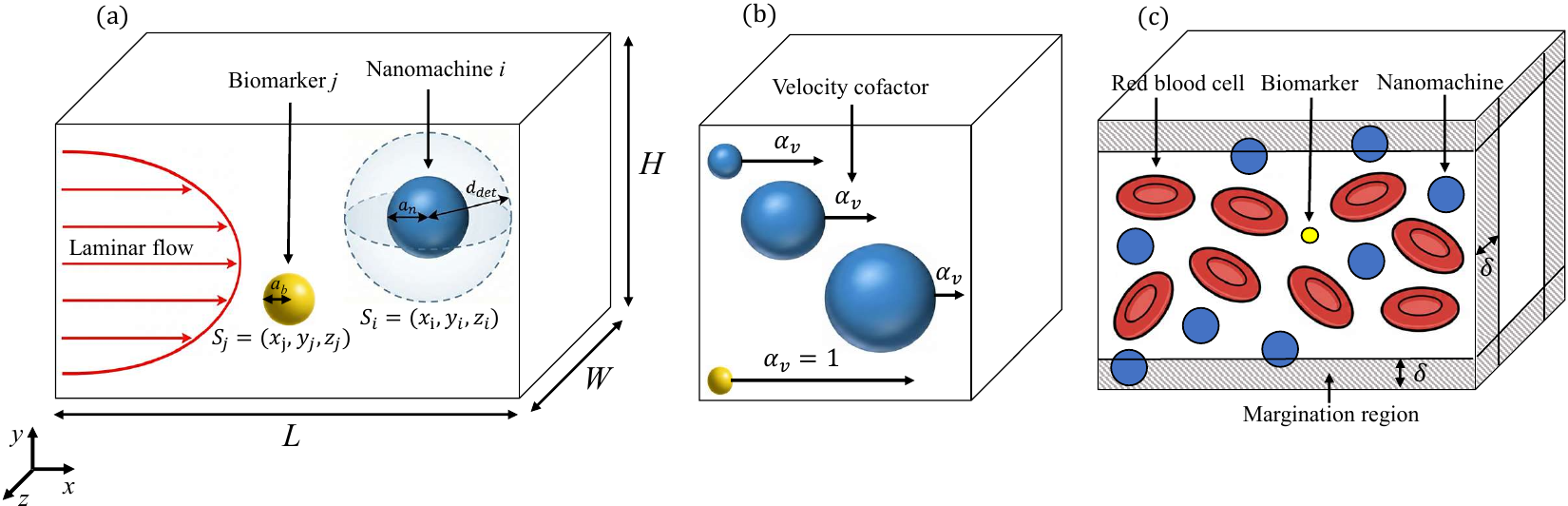}
\caption{Overview of the system model. (a) Biomarker detection mechanism in a generic vessel. (b) Nanomachine velocity cofactor assignment. (c) Modeling of nanomachine margination.}
\label{fig:General_Scenario}
\end{figure*}

\section{System Model}
\label{sec:model}
Fig. \ref{fig:General_Scenario} provides an overview of the system model considered in this paper, including general notation as well as the specific transport phenomena that are considered, all of which are further described in the next subsections.

\subsection{General Scenario}
In our scenario, we consider biomarkers and nanomachines flowing through human blood vessels. Biomarkers are protein molecules with a diameter in the order of 50~nm, released into the bloodstream by diseased or cancerous cells and widely used as indicators of pathological activity \cite{passaro2024cancer}. Nanomachines are engineered particles designed to detect these biomarkers directly within the vascular environment \cite{akyildiz2011nanonetworks}, and we consider them to take diameters in the range of 100~nm to 2~$\mu$m. In the human microcirculation, small blood vessels such as capillaries, venules, and arterioles exhibit an approximately cylindrical geometry over short spatial scales \cite{arciero2017mathematical}.

To model this environment, a three-dimensional simulation environment representing a segment of the microvascular network is considered. The simulation domain is modeled as a rectangular prism  with dimensions $L \times H \times W$, as depicted in Fig. \ref{fig:General_Scenario}(a), approximating the geometry of a small blood vessel and chosen for simulation simplicity. The values of $L$, $H$, and $W$ are chosen based on the typical dimensions of representative vessels including capillary, venule, and arterioles. The coordinate system is defined such that the longitudinal axis of the vessel aligns with the $x$-direction, while $y$ and $z$ correspond to the transverse and radial directions, respectively. 

Nanomachines and biomarkers are modeled as two distinct classes of particles with radii $a_n$ and $a_b$ circulating within the modeled vascular environment. In this context, let $\mathbf{s} = (x, y, z)$ denote the position of a particle within the domain. Nanomachine $i$ located at position $\mathbf{s}_i = (x_i, y_i, z_i)$ detects biomarker $j$ located at position $\mathbf{s}_j = (x_j, y_j, z_j)$ if their Euclidean distance satisfies
\begin{equation}
\| \mathbf{s}_i - \mathbf{s}_j \| \leq d_{\mathrm{det}},
\end{equation}
where $d_{\mathrm{det}}$ denotes the maximum detection range. 

\subsection{Flow Dynamics}
In microcirculation, blood flow is dominated by viscous forces and typically exhibits laminar behavior with structured velocity distributions \cite{hofmann2024molecular}. These characteristics strongly influence particle transport across the vessel.

To reflect these transport characteristics within a tractable modeling framework, flow dynamics are introduced progressively. A steady uniform flow is first considered as a simplifying assumption to obtain initial insight into system behavior. Uniform flow reduces model complexity and is commonly employed in MC studies to derive baseline performance trends \cite{wicke2018modeling, zheng2025system, noel2014diffusive}. To capture realistic vascular transport effects, laminar flow is subsequently incorporated into the vessel model. In this regime, the velocity follows a parabolic distribution, reaching its maximum at the centerline and approaching zero near the vessel walls. This behavior is modeled using the Hagen–Poiseuille equation,
\begin{equation} \label{eq:laminar}
v(r)=v_{\mathrm{max}}\left(1-\left(\frac{r}{R}\right)^2\right),
\end{equation}
where $v(r)$ denotes the flow velocity at a radial distance $r = \sqrt{x^2 + y^2}$ from the vessel centerline, $v_{\mathrm{max}}$ is the maximum velocity at the centerline, and $R$ is the vessel radius \cite{bruus2007theoretical}. The maximum velocity generally depends on the type of vessel that is considered.

\begin{figure*}[t]
    \centering
    \includegraphics[width=\linewidth]{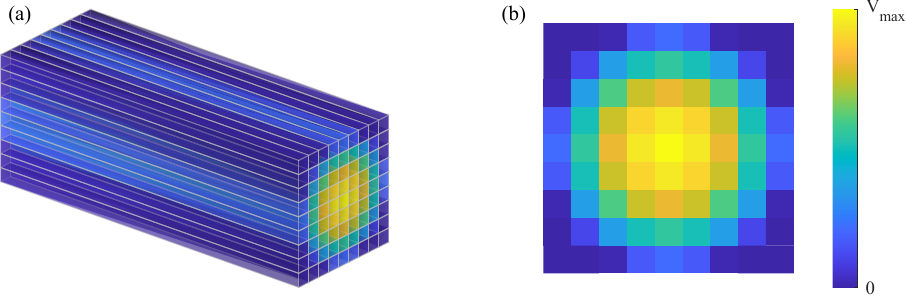}
    \caption{Laminar microvessel model. (a) Three-dimensional laminar velocity field in a microvessel. (b) Velocity heat map of vessel cross section.}
    \label{fig:Laminar microvessel modeling}
\end{figure*}

\subsection{Size-Dependent Particle Movement}

Particle transport in the microcirculation arises from the combined effects of Brownian motion and advection due to blood flow, with particle size strongly influencing how each mechanism contributes to overall movement \cite{liu2019unified}. 

Brownian motion, driven by thermal interactions between particles and the surrounding fluid, gives rise to diffusive transport. For spherical particles in microvascular environments characterized by low Reynolds numbers, the diffusion coefficient follows the Stokes--Einstein relation

\begin{equation}
D_{\mathrm{diff}} = \frac{k_B T}{6 \pi \mu a},
\label{eq:diffusion_coefficient}
\end{equation}
where $k_B$ is Boltzmann's constant, $T$ is the absolute temperature, $\mu$ represents the dynamic viscosity of blood, and $a$ is the particle radius \cite{bruus2007theoretical}.

In addition to diffusive transport driven by Brownian motion, particle motion in the microcirculation is strongly influenced by advection induced by blood flow. Smaller particles demonstrate faster transport, while larger counterparts tend to exhibit longer residence times and reduced effective transport under similar hemodynamic conditions \cite{xu2023size}. This behavior arises from more frequent interactions between larger particles and red blood cells, leading to prolonged residence time within the microvessel and diminished effective axial transport \cite{liu2019unified}.

To capture this behavior in a tractable manner, a velocity cofactor $0 \leq \alpha_v \leq 1$ is introduced such that the particle velocity at radial position $r$ is expressed as
\begin{equation}
v_{\text{particle}}(r) = \alpha_v v(r),
\label{eq:particle_velocity}
\end{equation}
where $v(r)$ denotes the local fluid velocity. A value of $\alpha_v = 0$ corresponds to nearly stationary particles, while $\alpha_v = 1$ indicates full advection with the surrounding flow. Biomarkers, owing to their nanoscale dimensions, are assumed to fully follow the blood flow ($\alpha_v = 1$). For nanomachines, the velocity cofactor is modeled as an inverse function of their radius, capturing the reduced effective advection and increased residence time experienced at larger sizes, as shown in Fig. \ref{fig:General_Scenario}(b). Accordingly, the velocity cofactor is defined as
\begin{equation}
\alpha_v = \frac{a_b}{a_n},
\label{eq:velocity_cofactor}
\end{equation}
where $a_n$ denotes the nanomachine radius and $a_b$ denotes the considered biomarker radius, serving as the normalization reference.

\subsection{Particle Margination}

As particles travel through the bloodstream, their motion is influenced by hydrodynamic interactions with red blood cells. This phenomenon, known as margination, causes particles to migrate laterally toward vessel walls, altering their spatial distribution within the flow \cite{liu2018nanoparticle}. Larger particles exhibit stronger lateral migration toward vessel walls under blood flow conditions due to increased hydrodynamic interactions \cite{liu2019unified, muller2014margination}. To analyze margination across different microvessel types and nanomachine sizes while considering velocity-dependent effects, a quantitative definition of margination is required. Based on the experimental findings in~\cite{carboni2016direct}, the near-wall region thickness $\delta$ is defined as $\delta = 0.1D$, where $D$ is the vessel diameter. The margination ratio $M$ is then expressed as
\begin{equation}
M = \frac{N_m}{N_{\text{total}}},
\label{eq:margination_ratio}
\end{equation}
where $N_m$ is the number of marginated particles located within the near-wall region and $N_{\text{total}}$ is the total number of particles within the vessel.

\section{Simulation Methodology}
\label{sec:methodology}

This section describes the simulation framework, including details on how the considered phenomena are implemented, value ranges for the simulation parameters, and the performance metrics that are calculated.

\subsection{Evaluation Framework}

The proposed system is evaluated through simulations conducted in AcCoRD, an open-source simulation framework designed for modeling molecular communication environments \cite{noel2017simulating}. AcCoRD is configured to represent physiologically realistic vascular conditions, incorporating flow, diffusion, reaction kinetics, and emitter–receiver interaction dynamics. However, modifications are needed to be made to include laminar flow, particle margination, and size-dependent propagation.

AcCoRD uses prisms as the basic building blocks of the simulated environment. Accordingly, a rectangular prism region is used to model a segment of the three-dimensional blood vessel environment. Using these environments as the transport medium, nanomachines are considered to be passively transported by the bloodstream. They do not possess self-propulsion capabilities and follow stochastic Brownian motion superimposed with advective flow. They are assumed to have adequate energy to perform sensing and communication functions throughout the entire simulation period.

At the start of the simulation, biomarkers and nanomachines are independently released from three predefined locations. Release points corresponding to the same particle type are separated by 15--20~$\mu$m, whereas the separation between biomarker and nanomachine release locations is 10--15~$\mu$m.

To implement laminar flow, we divided the blood vessel prism into multiple sub-prisms. Specifically, 81 sub-prisms are used for capillaries, 200 for venules, and 300 for arterioles. The number of sub-prisms is scaled with vessel diameter to provide finer spatial resolution for different vessel diameters and to accurately approximate the parabolic velocity profile across the vessel. The flow velocity of each sub-prism is then modified by calculating the distance of its center to the center of the vessel and using ~\eqref{eq:laminar}. Fig.~\ref{fig:Laminar microvessel modeling} illustrates a laminar microvessel model, in which the main prism is divided into multiple sub-prisms to represent laminar flow.

To approximate size-dependent particle transport, we defined multiple particle types representing biomarkers and nanomachines and used their size to define their flow in AcCoRD. In particular, we used $v_{max}$ from ~\eqref{eq:laminar} to determine the speed of biomarkers and then used ~\eqref{eq:particle_velocity} to obtain the nanomachine speed.

For modeling margination, two regions are defined for the release of nanomachines, as illustrated in Fig.~\ref{fig:General_Scenario}(c). One region spans the entire vessel cross section, while the other is restricted to the near wall region comprising the 10\% of the vessel space closest to the vessel boundary. Using the parameterized value of margination ratio $M$ from ~\eqref{eq:margination_ratio}, we select the amount of nanomachines to be released in each region. Biomarkers, on the other hand, are assumed to be distributed uniformly due to their small size and hence less tendency to be pushed towards the vessel walls.

\subsection{Simulation Parameters}
The simulation relies on a set of physiological and numerical parameters governing vessel geometry, blood flow characteristics, simulation duration, and particle transport behavior. Table \ref{tab:vessel_ranges} summarizes the vessel-specific configurations, reporting diameter and length together with the corresponding range of peak flow velocity for each microvascular segment. The vessel length is chosen as ten times the vessel diameter, providing an adequate axial domain to model fully developed laminar flow and particle transport, while maintaining computational efficiency. In line with the increase of blood velocity with vessel diameter, representative peak velocities of 1 mm/s, 2 mm/s, and 3 mm/s are selected for capillaries, venules, and arterioles, respectively \cite{bozuyuk2023microrobotic}. 

\begin{table}[!t]
\caption{Dimensions and Maximum Velocity of Microvessels \cite{bozuyuk2023microrobotic, lee2023internet}}
\centering
\renewcommand{\arraystretch}{1.05}
\setlength{\tabcolsep}{4pt}
\begin{tabular}{lcccc}
\hline
Parameter & Variable & Capillary & Venule & Arteriole \\ \hline
Diameter & $D$ & 9 $\upmu$m & 20 $\upmu$m & 30 $\upmu$m  \\
Length & $L$ & 90 $\upmu$m & 200 $\upmu$m & 300 $\upmu$m \\
Maximum Velocity & $v_{\max}$ & 0.5--1.5 mm/s & 1--3 mm/s & 1--100 mm/s \\ \hline
\end{tabular}  
\label{tab:vessel_ranges}
\end{table}

Table \ref{tab:particle_params} summarizes the particle dimensions and physical parameters used to characterize nanomachine and biomarker transport in the simulation framework. Nanomachine size and margination ratio are deliberately treated as open parameters to study their impact on the system performance, with default values set to 1 $\upmu$m and 0 (no margination), respectively. Dynamic viscosity and temperature are given by the usual conditions of blood in the human circulatory system. For the values of biomarkers, we selected exosomal carriers that reflect the characteristics of early-stage cancer indicators. In particular, ALPPL2$^{+}$ exosomes with a diameter of approximately 50 nm are assumed. Their concentration in the referenced measurements ranges from healthy controls with a reported value of $1.4 \times 10^{7}\,\mathrm{mL}^{-1}$, to NCC with a concentration of $5.3 \times 10^{8}\,\mathrm{mL}^{-1}$, and increases substantially to $2.0 \times 10^{9}\,\mathrm{mL}^{-1}$ in pancreatic cancer patients \cite{halder2025exosomal}. The NCC regime is adopted in this study, yielding approximately 1--10 biomarkers within the simulated vessel volumes. To ensure a uniform evaluation across all vessel types, the number of considered biomarkers is held constant at three.

\begin{table}[!t]
\caption{Particle Characteristics and System Parameters}
\centering
\renewcommand{\arraystretch}{1.1}
\setlength{\tabcolsep}{4pt}
\begin{tabular}{lccc}
\hline
Parameter & Variable & Value & Reference \\ \hline
Biomarker radius & $a_b$ & 25 nm & \cite{halder2025exosomal} \\
Nanomachine radius & $a_n$ & {100--2000} nm & \cite{liu2019unified, muller2014margination} \\
Margination ratio & $M$ & {0.05--0.60} & \cite{liu2019unified, muller2014margination} \\
Number of nanomachines & $N$ & {20--20000} & [—] \\
Temperature & $T$ & 310 K & [—] \\
Dynamic viscosity & $\mu$ & $4 \cdot 10^{-3}$ Pa$\cdot$s & \cite{froese2022flow} \\ \hline
\end{tabular}
\label{tab:particle_params}
\end{table}

Using \eqref{eq:diffusion_coefficient} and the parameters listed in the tables above, the diffusion coefficient of the considered biomarker is approximately $2.3 \times 10^{-12}\,\text{m}^2/\text{s}$. For nanomachines, the diffusion coefficient varies with particle size and spans from approximately $5.7 \times 10^{-14}\,\text{m}^2/\text{s}$ for the largest considered radius to $1.1 \times 10^{-12}\,\text{m}^2/\text{s}$ for the smallest considered radius. 

In addition to diffusion, the contribution of advective transport is quantified through the velocity cofactor defined in  \eqref{eq:velocity_cofactor}. Using the considered biomarker radius as the reference scale, nanomachine velocity cofactors decrease from 0.25 to 0.0125 for nanomachine radii spanning 100--2000~nm, reflecting the progressive reduction in effective axial transport with increasing size.

\subsection{Performance Metrics}
\label{subsec:performance_metric}
We next define the metric used to assess nanomachine detection performance. Detection probability $P_d$ is defined as the ratio of detected biomarkers to the total number of biomarkers present in the simulated environment,
\begin{equation}
P_d = \frac{N_{\mathrm{bio,det}}}{N_{\mathrm{bio,tot}}},
\end{equation}
where $N_{\mathrm{bio,det}}$ is the number of \textit{biomarkers} successfully detected by nanomachines during the observation period, and $N_{\mathrm{bio,tot}}$ is the total number of biomarkers introduced in the simulation. This metric is averaged over multiple simulation trials to ensure statistical reliability. 

\section{Results}
\label{sec:results}

We first investigate the impact of flow regime and vessel geometry on detection probability. Simulations are conducted under both uniform and laminar flow conditions. For each flow type, detection probability is evaluated as a function of number of nanomachines in capillary, venule, and arteriole configurations.

\begin{figure*}[t]
    \centering
    \includegraphics[width=\textwidth]{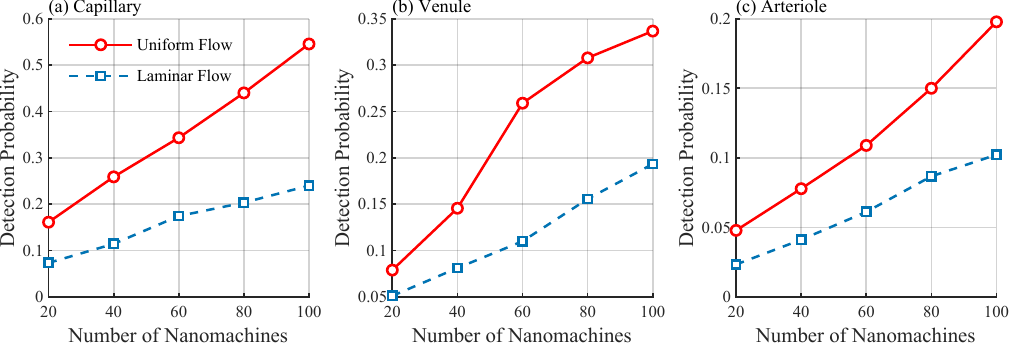}
    \caption{Detection probability versus number of nanomachines in (a) capillary, (b) venule, and (c) arteriole under uniform and laminar flow.}
    \label{fig:dp_all vessels_uniform_and_laminar_comparison}
\end{figure*}

Fig.~\ref{fig:dp_all vessels_uniform_and_laminar_comparison} summarizes detection probability versus nanomachine count for capillaries, venules, and arterioles under uniform and laminar flow conditions. A nanomachine radius of 1 $\upmu$m is assumed across all vessel types, yielding a very low velocity cofactor ($\alpha_v = 0.025$), effectively approaching zero and reflecting minimal advective transport. The results show that increasing the number of nanomachines consistently improves detection probability under both flow models, while laminar flow consistently results in lower detection compared to uniform flow due to near-wall velocity attenuation. Nevertheless, laminar flow is adopted as a physiologically realistic representation of microvascular transport. Moreover, detection probability decreases systematically with increasing vessel dimensions for both flow conditions. Transitioning from a capillary to an arteriole, which is approximately three times larger in diameter and length, leads to more than a twofold reduction in detection probability when the number of nanomachines is fixed.

\begin{figure}[t]
    \centering
    \includegraphics[width=\columnwidth]{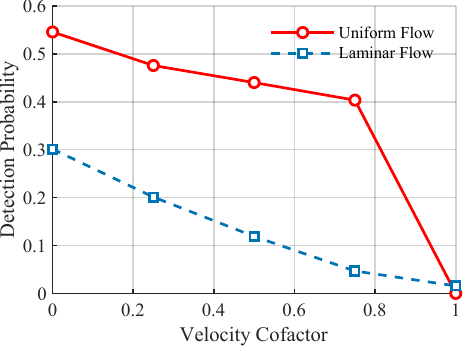}
    \caption{Detection probability as a function of nanomachine velocity cofactor in capillary under uniform and laminar flow, with $N$ fixed at 100.}
    \label{fig:capillary_dp_comparison_uniform_laminar}
\end{figure}

\begin{figure}[t]
    \centering
    \includegraphics[width=\columnwidth]{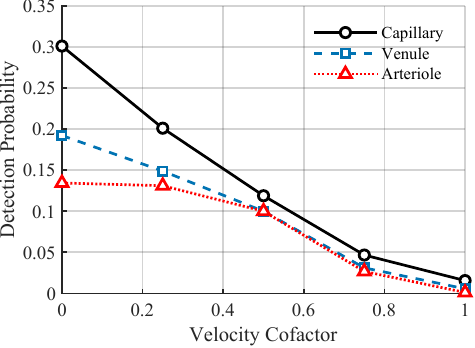}
    \caption{Detection probability as a function of nanomachine velocity cofactor in capillary, venule, and arteriole under laminar flow, with $N$ fixed at 100.}
    \label{fig:velocity_cofactor_all_vessels}
\end{figure}

Next, the role of the velocity cofactor in detection performance is examined, with the nanomachine radius fixed at 1~$\mu$m. Figs.~\ref{fig:capillary_dp_comparison_uniform_laminar} and~\ref{fig:velocity_cofactor_all_vessels} collectively demonstrate that increasing the velocity cofactor degrades detection performance for a fixed number of 100 nanomachines. Fig.~\ref{fig:capillary_dp_comparison_uniform_laminar} compares uniform and laminar flow models for capillary, showing that under laminar flow the detection probability decreases smoothly as the velocity cofactor increases, reflecting the realistic influence of spatially varying flow velocities. Uniform flow, by comparison, exhibits a sharper degradation in detection at higher velocity cofactors, suggesting an overestimation of advective effects. Fig.~\ref{fig:velocity_cofactor_all_vessels} generalizes this behavior to capillary, venule, and arteriole under laminar flow conditions. For all vessel types, the detection probability reaches its minimum at a velocity cofactor of 1, representing the limiting case in which nanomachines and biomarkers travel at the same velocity and relative motion is minimized.

\begin{figure*}[t]
    \centering
    \includegraphics[width=\textwidth]{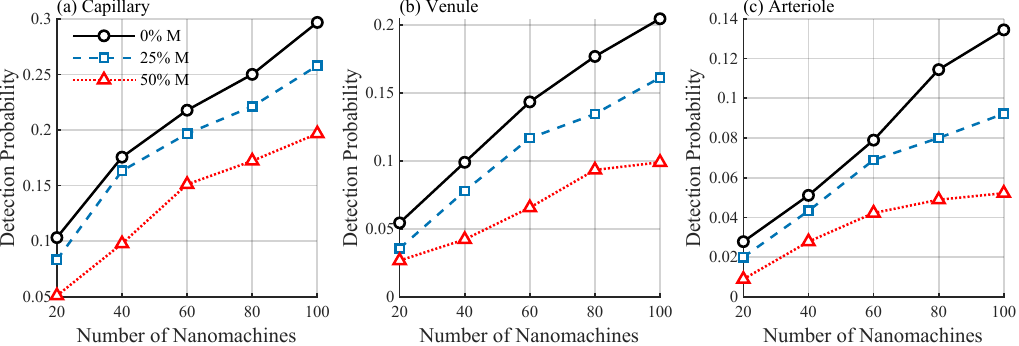}
    \caption{Detection probability versus number of nanomachines in (a) capillary, (b) venule, and (c) arteriole at different margination ratios under laminar flow.}
    \label{fig:dp_all_vessels_margination_comparison}
\end{figure*}

\begin{figure*}[t]
    \centering
    \includegraphics[width=\textwidth]{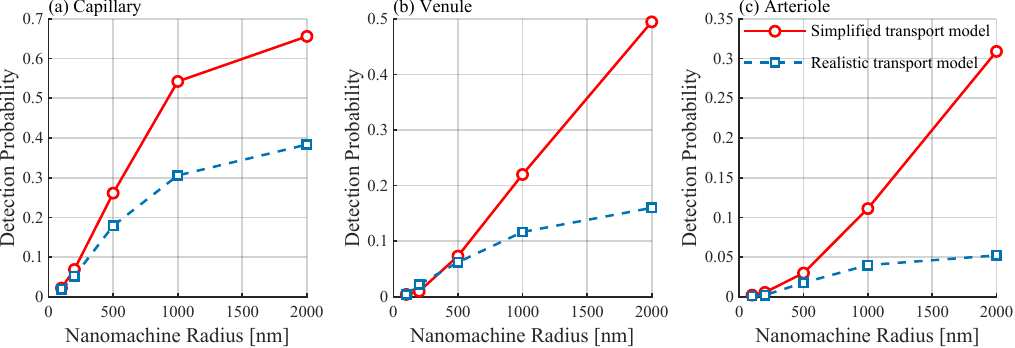}
    \caption{Detection probability as a function of nanomachine size in (a) capillary, (b) venule, and (c) arteriole, with $N$ fixed at 100.}
    \label{fig:dp_nanomachine_size_system_level}
\end{figure*}

Thereafter, we analyze the impact of margination on detection performance. In this scenario, the nanomachine radius is fixed at 1~$\mu$m across all vessel types, resulting in a very low velocity cofactor ($\alpha_v = 0.025$). As illustrated in Fig.~\ref{fig:dp_all_vessels_margination_comparison}, an increase in the margination ratio results in a decrease in detection probability across all vessel types. With the number of nanomachines fixed at 100, the detection probability at 0\% margination is approximately 1.5 times higher than at 50\% margination in capillaries, and about two times higher in venules and arterioles.

Subsequently, the effect of nanomachine size on detection probability is investigated while keeping the number of nanomachines fixed at 100. Fig.~\ref{fig:dp_nanomachine_size_system_level} compares a simplified transport model employing uniform flow, in which margination effects and velocity cofactor modeling are not considered, with a more realistic transport model that incorporates laminar flow, margination effects, and velocity cofactor modeling. In the realistic model, margination is parameterized based on nanomachine size, with values ranging from 0.05 for the smallest nanomachines to 0.60 for the largest.

Across capillaries, venules, and arterioles, detection probability increases with nanomachine size, reflecting the larger effective sensing area of bigger nanomachines. However, this increase is substantially slower in the realistic model due to the adoption of laminar flow, which reduces effective transport near the vessel walls compared to the simplified model. For nanomachine radii exceeding 1~µm, margination effects become increasingly significant, resulting in a pronounced gap in the detection probability between the two models.

Among vessel types, capillaries consistently provide the highest detection efficiency across all sizes. For example, at a nanomachine radius of 2~µm, capillary detection probability exceeds that of venules by more than a factor of two and that of arterioles by over a factor of seven, while venules outperform arterioles by approximately a factor of three. This trend is primarily driven by the smaller capillary diameter, which confines particle dispersion and increases interaction likelihood compared to wider vessels.

Tables~III--V provide a design-oriented summary of the simulation results, indicating that achieving higher detection probabilities requires an increased number of nanomachines across all vessel types and nanomachine sizes. The results further show that larger nanomachine sizes require fewer nanomachines to attain the target detection probabilities compared to smaller sizes. In addition, capillary achieves the target detection probabilities with a significantly smaller number of nanomachines than venule and arteriole.

\begin{table}[t]
\caption{Estimated Nanomachines for Target Detection Probabilities}
\label{tab:combined_vessels}
\centering
\begin{tabular}{c c c c}
\hline
Vessel type & $a_n$ & $N(P_d = 0.25)$ & $N(P_d = 0.5)$ \\
\hline
          & 500\,nm  & $170$ & $567$ \\
Capillary & 1000\,nm & $75$ & $250$ \\
          & 2000\,nm & $55$ & $180$ \\
\hline
          & 500\,nm  & $500$ & $1800$ \\
Venule    & 1000\,nm & $222$ & $850$ \\
          & 2000\,nm & $170$ & $620$ \\
\hline
          & 500\,nm  & $2500$ & $20000$ \\
Arteriole & 1000\,nm & $1250$ & $10000$ \\
          & 2000\,nm & $1000$ & $5000$ \\
\hline
\end{tabular}
\end{table}

\section{Conclusion}
\label{sec:conclusion}

This paper presents a systematic evaluation of biomarker detection probability for early-stage cancer across capillary, venular, and arteriolar vessels by comparing a simplified model with more realistic transport mechanisms. The results show that incorporating realistic transport mechanisms leads to a substantial reduction in detection probability across all considered vessel types compared to the simplified model. Among them, capillary consistently provides the highest detection efficiency. The insights gained from this study can guide the design and evaluation of future nanomachine-enabled detection strategies in the microcirculation. These results provide a foundation for studying biomarker detection in branching vascular structures, such as bifurcations and junctions, in future investigations.

\ifCLASSOPTIONcaptionsoff
  \newpage
\fi



%

\bibliographystyle{IEEEtran}
\bibliography{refs}




%









\end{document}